\ifx\mnmacrosloaded\undefined \input mn\fi

\pageoffset{-2.5pc}{0pc}
\overfullrule=0pt


\begintopmatter
\title{Gamma-ray bursts from internal shocks \hfil\break in a relativistic wind:\hfil\break temporal and spectral properties}
\author{F. Daigne and R. Mochkovitch}
\affiliation{Institut d'Astrophysique de Paris, CNRS, 98 bis Boulevard Arago, 75014 Paris, France}
\shortauthor{F. Daigne and R.Mochkovitch}
\shorttitle{Gamma-ray bursts from internal shocks in a relativistic wind}

\abstract{We construct models for gamma-ray bursts where 
the emission comes from internal shocks in a relativistic wind with a
highly non uniform distribution of the Lorentz factor. We follow
the evolution of the wind using a very simplified approach where a large 
number of layers interact by direct collisions but where all pressure 
waves have been suppressed. We suppose that the magnetic field and the 
electron Lorentz factor reach large equipartition values in the shocks.
Synchrotron photons emitted by the relativistic electrons have a typical
energy in the gamma-ray range in the observer frame.
Synthetic bursts are constructed as the sum of the contributions from all the
internal elementary shocks and their temporal and spectral properties are
compared to the observations. We reproduce the diversity of burst profiles,
the ``FRED'' shape of individual pulses and the short time scale
variability. Synthetic bursts also satisfy the duration-hardness relation
and individual pulses are found to be narrower at high energy, in
agreement with the observations. These results suggest that 
internal shocks in a relativistic wind may indeed be at the origin of 
gamma-ray bursts. A potential
problem is however the relatively low efficiency of the dissipation process.
If the relativistic wind is powered by accretion from a disc to a stellar mass
black hole it implies that a substantial fraction of the available energy is
injected into the wind.}


\pagerange{1--12}
\pubyear{1997}
\maketitle


%
%

\section{Introduction}
Since 1991 the BATSE experiment on board the Compton GRO satellite has observed
more than 1800 gamma-ray bursts (hereafter GRBs). The burst distribution is 
isotropic over the sky but non homogeneous in distance (Fishman and Meegan, 
1995 and references therein) which has been
regarded as a strong indication that GRBs lie 
at cosmological distances (Paczy\'nski 1991). However the possibility that GRBs belong to 
a large galactic halo (Hartmann et al. 1994) could not be excluded a priori.
This long standing controversy (see Nemiroff et al. 1995) about the burst distance scale may finally be solved
by the recent observations of transient optical counterparts for two GRBs.
In the case of GRB 970228 
(van Paradijs et al. 1997; Sahu et al. 1997a) 
the point-like counterpart appears to be 
associated to an extended source which is probably a distant galaxy.
The case of GRB 970508 is even more spectacular since the 
spectrum of the counterpart shows
Fe II and Mg II lines due to absorbers on the line of sight at a 
redshift $z=0.835$ (Metz\-ger et al. 1997). If the association of the visible source to the
burst is confirmed, GRB 970508 must be a very distant object at 
$z\ge 0.835$. 
If GRBs are placed at cosmo\-lo\-gi\-cal distances 
the Log$N$ -- Log$P$ (peak
flux) curve and the value of $<V/V_{\rm max}>$ can be naturally 
interpreted in terms 
of cosmological effects and indicate that GRBs have 
typical redshifts in the range 0.3 -- 1 (Piran 1992; Mao \& Paczy\'nski 1992;
Fenimore et al. 1993).
\par
Modelling GRBs is a difficult task
due to the extreme diversity
of burst profiles, the non thermal spectra and the lack of 
clear signature for the emission processes involved. Most cosmological models 
however share some common characte\-ris\-tics. The source, which must be able 
to release (between 10 keV and 10 MeV) an energy 
$E_{\gamma}\ga {10^{51}\over 4\pi}$ erg.sr$^{-1}$
on a time scale of seconds, is generally
supposed to be a stellar mass black hole accreting material from a 
disc. Such a configuration can result from the coalescence of two neutron 
stars (Eichler et al. 1989; Paczy\'nski 1991; Narayan, Paczy\'nski \& Piran 
1992), 
the disruption of the neutron star in a neutron star -- black
hole binary (Narayan Paczy\'nski \& Piran 1992; Mochkovitch et al. 1993) 
or the collapse of a massive star (Woosley 1993).
The power emitted by cosmological GRBs is orders of magnitude larger than the Eddington
luminosity and cannot come directly from the disc surface. The 
released energy instead drives a wind which has to become relativistic
both to produce gamma-rays and to avoid photon-photon annihilation along the line
of sight (Baring 1995; Sari \& Piran 1997a). The Lorentz factor $\Gamma$ must reach values of $10^2$ --
$10^3$ which limits the allowed amount of baryonic pollution in the flow to a
very low level. A few mechanisms which could possibly achieve such a severe
constraint have been proposed: ({\it i}) magnetically driven outflow 
originating
from the disc or powered by the Blandford-Znajek (1977) effect
(Thompson 1994; M\'esz\'aros \& Rees 1997a),
{\it ii}) reconnection of 
magnetic field lines in the disc corona (Narayan, Paczy\'nski \& Piran 1992) 
or ({\it iii}) 
neutrino-antineutrino annihilation 
in a funnel along the rotation axis of the system (M\'esz\'aros \& Rees 1992;
Mochkovitch et al. 1993, 1995). 
It is supposed in ({\it i}) and ({\it ii}) that the field 
has reached huge values $B \ga 10^{15}$ G.
Concerning ({\it iii}) Ruffert et al. (1997) have shown recently that $\nu{\bar \nu}$
annihilation does not provide enough energy to account for cosmological GRBs
except may be for very massive discs, as those which could result from
the collapse of a massive star.\par
The energy initially stored in kinetic form within the relativistic wind
must then be converted into gamma-rays. This can be done during the
deceleration of the wind resul\-ting from its interaction with the
interstellar medium (Rees \& M\'esz\'aros 1992; M\'esz\'aros \& Rees 1993) or a dense radiation field
(Shemi 1994). In the first case the
emission comes from electrons accelerated in the forward and reverse 
shocks which then radiate synchrotron and inverse Compton photons in a 
magnetic field frozen in the wind or which has come 
to equipartition with the shocked material (M\'esz\'aros, Laguna \& Rees 1993).
In the second case, photons of the ambient radiation field (in the central
regions of globular clusters or active galactic nuclei) interact with the
electrons (Shemi 1994; Shaviv \& Dar 1995) or ions (Shaviv \& Dar 1996) of the wind and undergo a boost in energy 
by a factor $\Gamma^2$. With $\Gamma\la 10^3$, optical photons can be
shifted to the gamma-ray range. In these two models the duration of the burst
is $t_{\rm b}\sim {r_{\rm dec}\over c \Gamma^2}$ where $r_{\rm dec}$ is the deceleration radius of the
wind. Another possibility suggested by Rees \& M\'esz\'aros (1994) consists to suppose that
the Lorentz factor in the wind is variable so that successive shells
can have large relative velocities leading to the formation of internal shocks.
The energy which is dissipated in these shocks can then be radiated as gamma-rays
via the production of pions in proton-proton collisions (Paczy\'nski \& Xu
1994) or by 
synchrotron (or inverse Compton) emission of accelerated electrons. 
There are several important differences between the deceleration and internal
shock models. In deceleration models, the duration and time profile of the burst
depend on $\Gamma$, on the value of the deceleration radius and 
on the structure of the emitting shell. In internal shock models,
the duration of the burst is directly related to the duration of energy injection 
at the source and the time profile is essentially determined by the 
variations of the Lorentz factor (Sari \& Piran 1997a). 
\par
Burst
profiles and spectra have already been obtained (sometimes in a rather detailed 
manner) in the case of deceleration models (Fenimore, 
Madras \& Nayakshin 1997; Panaitescu et al. 1997; Panaitescu \& M\'esz\'aros 
1997) and
the purpose of this work
is to present the same kind of quantitative analysis for 
internal shock models.
In Sect. 2 we describe our method to follow the evolution of the relativistic wind
and we discuss the emission processes; Sect. 3 and 4 respectively deal 
with the temporal and spectral properties of our synthetic burst models and 
Sect. 5 is 
the conclusion. 

%
%

\section{A simple model of the relativistic wind}
\subsection{Description of the model}
We do not discuss in this paper the nature of the source (coalescence of two 
neutron stars, neutron star -- black hole binary, collapse of a massive 
star or something
else) which is initially responsible for the energy release. We also 
suppose that a relativistic wind carrying the energy has emerged from the 
source, with an average Lorentz factor ${\bar \Gamma}$ of a few hundreds.
\par
To study the evolution of the relativistic wind we have developed a very 
simple model 
where a succession of layers are emitted 
every 2 ms with a varying Lorentz factor, during a total time 
$t_{\rm w}$. The mass of the layers 
is proportional to $1/\Gamma$ so that the energy injection rate 
is constant. We follow the layers as the wind expands and when a rapid layer
(of mass $m_1$ and Lorentz factor $\Gamma_1$)
catches up with a slower one ($m_2$, $\Gamma_2<\Gamma_1$) they collide and merge to form a single shell of resulting
Lorentz factor $\Gamma_{\rm r}$.
If the dissipated energy
$$e=m_1 c^2 \Gamma_1+m_2 c^2 \Gamma_2-(m_1+m_2) c^2 \Gamma_{\rm r}\;,\eqno\stepeq$$
can be radiated in the gamma-ray range on a time scale shorter than
the shell expansion time (in the comoving frame)
$$t_{\rm ex}\simeq{r\over c \Gamma_{\rm r}}\;,\eqno\stepeq$$ 
where $r$ is the shell radius, the burst profile will be made by a succession
of elementary contributions of duration (in the observer frame)
$$\Delta t\simeq{r\over 2 c \Gamma^2_{\rm r}}\;.\eqno\stepeq$$ 
We estimate $\Gamma_{\rm r}$ by considering
that most of the energy available in the collision has been already 
released when the less massive of the two layers has swept up a mass
comparable to its own mass in the other layer (internal shocks being only
mildly relativistic).
Then
$$\Gamma_{\rm r}\simeq \sqrt{\Gamma_1 \Gamma_2}\;,\eqno\stepeq$$ 
and
$$e=mc^2(\Gamma_1+\Gamma_2-2 \Gamma_{\rm r})\hskip 0.5truecm,\hskip 0.5truecm
m={\rm min}(m_1,m_2)\;.\eqno\stepeq$$ 
After the complete redistribution of momentum and energy the Lorentz factor
of the merged layer finally becomes
$$\Gamma_{\rm f}=\sqrt{\Gamma_1\Gamma_2{m_1\Gamma_1+m_2\Gamma_2\over
m_1\Gamma_2+m_2\Gamma_1}}\;.\eqno\stepeq$$ 
Two more conditions have to be satisfied for an elementary shock to be produced and 
observed: {\it i}) the two layers must collide at a relative
velocity larger than the local sound speed and {\it ii})
the wind must be transparent to the emitted photons. The relative velocity
of the two layers is given by
$${v_{\rm rel}\over c}\simeq {\Gamma_1^2-\Gamma_2^2\over \Gamma_1^2+\Gamma_2^2}
\;,\eqno\stepeq$$ 
with $\Gamma_1>\Gamma_2$ and $\Gamma_{1,2}\gg 1$. We have adopted
a sound speed $v_{\rm s}/c=0.1$ 
but we have checked that other choices make little differences in the
results since the main contribution to the burst comes from shocks 
corresponding to large differences of the Lorentz factor ${\Gamma_1\over
\Gamma_2}\ga 2$. \par
The transparency of the wind to the emitted photons has been computed in
the following way: in each collision between layers a ``photon shell'' is 
generated which then catches up with all the layers ahead of it. It travels
through a total optical depth
$$\tau=\kappa_{\rm T}\sum_{i>i_{\rm m}} {m_i\over 4\pi r_i^2}\;,\eqno\stepeq$$ 
where $\kappa_{\rm T}$ is the Thomson opacity and $r_i$ the radius of layer $i$
when it is reached by the photon shell. The sum is over all
indices $i$ larger than $i_{\rm m}$, which corresponds to the two merged layers.
For the average Lorentz factor $\bar{\Gamma}\ga 100$ used here it appears
that the wind is transparent to the emitted photons except for a few
early collisions.
\par
The arrival time of each of the elementary contributions from internal 
shocks is calculated relatively
to a signal which would have travelled at the speed of light from the
source to the observer. It is given by
$$t_{\rm a}=t_{\rm e}-{r\over c}\;,\eqno\stepeq$$ 
where $t_{\rm e}$ is the emission time and $r$ the distance to the source 
of the two merged layers. The evolution of the system is 
followed until all the 
layers are ordered with $\Gamma$ decreasing from the front to the back of the
wind.
The efficiency of the
dissipation process can then be obtained as 
$$f_{\rm d}={\sum_{s} e_s\over\sum_{i} m_i c^2 \Gamma_i}\;,\eqno\stepeq$$ 
where the $e_s$ are the energies released in each of the internal elementary shocks and 
the $m_i$,
$\Gamma_i$ are the initial masses and Lorentz factors of the layers.
\par
This very simple approach is naturally very crude because it neglects all pressure
waves propagating throughout the layers. Nevertheless, we expect that it can
still capture the basic features of the real process. As a first example we
have represented in Fig.1 the evolution of the Lorentz factor at different
times when the initial distribution of $\Gamma$ consists of 5000 layers
(i.e. $t_{\rm w}=10$ s) with $\Gamma(n)=400$ for $n=1$ to 4000 and 
$\Gamma(n)=100$ 
for $n=4001$ to 5000
($n=1$ for the last emitted layer).
Such a wind is then divided between a ``slow'' part which is emitted first and a rapid part
which will progressively collide against the slow part. The total masses injected in the slow and rapid parts
are equal. After the first collision has 
occurred a separation layer of increasing mass 
is formed between the slow and rapid parts.
We have plotted in Fig.2 the values of 
$t_{\rm e}$, $\Gamma_{\rm r}$,
$\Delta t$ and 
$e$ for the elementary shocks as a function of arrival time $t_{\rm a}$. The evolution 
of the shell system is essentially completed after a time $t_{\rm e}\sim t_{\rm var} {\bar
\Gamma}^2$, where $t_{\rm var}$ is the characteristic time scale for 
the variations of the Lorentz 
factor. As expected, the plots of $\Gamma_{\rm r}$, $\Delta t$
and $e$
have two branches, corresponding to collisions taking place on both
sides of the separation layer. They mimic the forward and reverse shocks
which, in our simple model, are located on a single surface.\par
The efficiency $f_{\rm d}$ of the dissipation process is rather low, 10\% in this
specific case, typically less than 15\%. This is a severe problem,
which can become even worse if only a fraction of the dissipated energy 
is radiated in the gamma-ray range. This issue is examined in the next section.
\beginfigure{1}
\caption{{\bf Figure 1.} Distribution of the Lorentz factor in the wind at different times
in lagrangian coordinates $m/m_{\rm tot}$ ($m_{\rm tot}$ being the total mass
of the wind). The initial distribution ($t=0$) consists of 5000
layers with $\Gamma=100$ in the 1000 first emitted layers 
and $\Gamma=400$ in the
rest of the wind. Since the mass of the layers is proportional to $1/\Gamma$
the slow and rapid parts contain an equal mass. At $t=3.4\;10^4$ s
a forward and a reverse shock are propagating into the wind. In our simple model
they are located on a single eulerian shell of increasing mass. At
$t=1.7\;10^5$ s the forward shock has already crossed the whole slow part
and at $t=5.4\;10^5\; \rm s$ the reverse shock reaches the end of the rapid part.
The dashed line shows the initial distribution of the Lorentz factor.}
\endfigure
\beginfigure{2}
\caption{{\bf Figure 2.} Physical parameters of the elementary shocks. Upper panel:
(thick line) emission time $t_{\rm e}$ and (thin line) duration $\Delta t$ 
of the signal (in the observer frame)
both as a function of arrival time $t_{\rm a}$.
Lower panel: (thick line) energy $e$ dissipated in the shock in arbitrary unit
and 
(thin line) Lorentz factor $\Gamma_{\rm r}$ at the shock location.}
\endfigure

\subsection{Emission processes}
The process by which the dissipated energy is finally radiated
depends on the energy distribution of protons and electrons in the 
shocked material and on
the values of the comoving density and magnetic field.
The average energy which is dissipated per proton in a shock between two layers
of equal mass is given by 
$$
\displaylines{
\hbox to \hsize{}\cr
\epsilon=(\Gamma_{\rm int}-1)m_{\rm p} c^2\hfill \cr
{\rm with}\ \ \Gamma_{\rm int}={1\over 2}\left[\left({\Gamma_1\over \Gamma_2}\right)^{1/2}+\left({\Gamma_2\over \Gamma_1}\right)^{1/2}\right]\;,\hfill\stepeq\cr} 
$$
where $\Gamma_{\rm int}$ is the Lorentz factor for internal motions in the 
shocked material.
With $\Gamma_1/\Gamma_2=4$, which corresponds to a mildly relativistic
shock ($v_{\rm rel}/c=0.88$), $\epsilon\ga 200$ MeV. \par
A mechanism which could directly extract the energy from the protons
has been proposed by Paczy\'nski \& Xu (1994) since the value of 
$\epsilon$ is large
enough for pions production in pp collisions. The pions then decay 
with the emission of
gamma-rays. The global efficiency of this mechanism is low, 
of the order of $10^{-3}$
for $\Gamma_1/\Gamma_2=4$. 
\par
Another possibility, considered by Rees \& M\'esz\'aros (1994),
Papathanassiou \& M\'es\-z\'a\-ros (1996) and Sari \& Piran
(1997a),
consists to suppose that the electrons have come into (at least partial) 
equipartition with the protons. If a fraction $\alpha_{\rm e}$ of the 
dissipated energy
goes to the electrons their characteristic Lorentz factor will be
$$\Gamma_{\rm e}\simeq \alpha_{\rm e} {\epsilon\over m_{\rm e} c^2}\;,\eqno\stepeq$$ 
which, for $\alpha_{\rm e}=1/3$ (corresponding to a complete equipartition
between protons, electrons and the magnetic field) 
and $\Gamma_1/\Gamma_2=4$
yields $\Gamma_{\rm e}\sim 150$.
The equipartition magnetic field is given by
$$B_{\rm eq}\simeq \left(8\pi\alpha_{\rm B} n \epsilon\right)^{1/2}\;,\eqno\stepeq$$ 
where $\alpha_{\rm B}\la 1$ and $n$ is the comoving proton number density
$$n\simeq{\dot{M}\over 4\pi r^2 {\bar {\Gamma}} m_{\rm p} c}\simeq
     {\dot{E}\over 4\pi r^2 {\bar {\Gamma}}^2 m_{\rm p} c^3}\;.\eqno\stepeq$$ 
Assuming $\dot{E}=10^{52}$ erg.s$^{-1}$, $\bar{\Gamma}=300$, $t_{\rm var}=1$ s
and $\alpha_{\rm B}=1/3$
the equipartition magnetic field at a radius $r\sim c t_{\rm var}{\bar{\Gamma}}^2$
where most of the collisions take place is 
$B_{\rm {eq}} \sim (10^2 - 10^3)\; \rm G$ depending on the ratio $\Gamma_1/\Gamma_2$.
\par
Synchrotron emission
by the accelerated electrons in the magnetic field occurs at a typical energy
(in the observer frame)
$$E_{\rm syn}=50 \left({\Gamma_{\rm r}\over 300}\right)
\left({B\over 1000 {\rm G}}\right)\left({\Gamma_{\rm e}\over 100}\right)^2\ \ 
{\rm eV}\;.\eqno\stepeq$$ 
which corresponds to the UV range for $\Gamma_{\rm e}=100$. Gamma-rays 
can be produced by inverse Compton scattering on the synchrotron photons. Then
$$E_{\rm IC}\simeq E_{\rm syn} \Gamma_{\rm e}^2=500
\left({\Gamma_{\rm r}\over 300}\right)
\left({B\over 1000 {\rm G}}\right)\left({\Gamma_{\rm e}\over 100}\right)^4\ \ 
{\rm keV}\;,\eqno\stepeq$$ 
and the fraction of the total power which is radiated by the inverse Compton process 
is
$$\alpha_{\rm IC}={\tau_{\star} \Gamma_{\rm e}^2\over 1+\tau_{\star} 
\Gamma_{\rm e}^2}\;,\eqno\stepeq$$ 
where
$$\tau_{\star}={\kappa_{\rm T} M_{\star}\over 4\pi r_{\star}^2}\;,\eqno\stepeq$$ 
is the optical depth of the shell of mass $M_{\star}$ and radius $r_{\star}$
which contains the relativistic electrons. An estimate of $M_{\star}$ is 
$$M_{\star}={t_{\rm syn}\over 1+ \tau_{\star}\Gamma_{\rm e}^2}
{\dot M}_{\rm shock}\;,\eqno\stepeq$$ 
where
$$t_{\rm syn}=6 \left({\Gamma_{\rm e}\over 100}\right)^{-1}
\left({B\over 1000 {\rm G}}
\right)^{-2}\ \ {\rm s}\;,\eqno\stepeq$$ 
is the synchrotron time of the relativistic electrons and 
$\dot{M}_{\rm shock}$ the mass flow rate across the shock, 
both in the comoving frame of the shocked material.
Since the shock moves with a Lorentz factor 
$\Gamma_{\rm s}\simeq {\bar \Gamma}$,
${\dot M}_{\rm shock}$ can be approximated by
$$\dot{M}_{\rm shock}\simeq {\dot{M}\over \bar{\Gamma}}\;.\eqno\stepeq$$ 
From (18), (19) and (21) we obtain an implicit expression for 
$\tau_{\star}\Gamma_{\rm e}^2$
$$\tau_{\star}\Gamma_{\rm e}^2={\kappa_{\rm T}\dot{M}_{\rm shock} 
t_{\rm syn} \Gamma_{\rm e}^2\over 4\pi r_{\star}^2
(1+\tau_{\star}\Gamma_{\rm e}^2)}
\;.\eqno\stepeq$$ 
With $r_{\star}\sim c t_{\rm var}\bar{\Gamma}^2$, relations (20) and (21)
for $t_{\rm syn}$ and $\dot{M}_{\rm shock}$, (13) and (14) for 
$B_{\rm{eq}}$ and $n$, Eq. 22 gives (for $\kappa_{\rm T}= 0.2$
cm$^2$.g$^{-1}$) 
$$\tau_{\star}\Gamma_{\rm e}^2 (1+\tau_{\star}\Gamma_{\rm e}^2)
\simeq 0.3 {\alpha_{\rm e}\over \alpha_{\rm B}}\;,\eqno\stepeq$$ 
which, for $\alpha_{\rm e}\sim \alpha_{\rm B}$ yields 
$\tau_{\star}\Gamma_{\rm e}^2\simeq 0.24$ and $\alpha_{\rm IC}\simeq 0.19$.
A larger fraction of the dissipated energy can be converted to gamma-rays
if the magnetic field does not reach equipartition. With for
example ${\alpha_{\rm e}\over \alpha_{\rm B}}=100$, 
$\tau_{\star}\Gamma_{\rm e}^2$ rises to 5 and $\alpha_{\rm IC}$ to 0.83.
Smaller values of $\alpha_{\rm B}$ could still increase $\alpha_{\rm IC}$
but the energy of the inverse Compton photons would then become too
small.\par
Conversely, if the electron Lorentz factor is large enough gamma-rays 
can be directly produced by synchrotron emission. This will be the case if,
instead of (12) one uses the expression of $\Gamma_{\rm e}$ given by
Bykov \& M\'esz\'aros (1996) who
consider the scattering of electrons
by turbulent magnetic field fluctuations. They get
$$\Gamma_{\rm e}\sim \left[\left({\alpha_{\rm M} \over \zeta}\right)\left(
{\epsilon\over m_{\rm e} c^2}\right)\right]^{1/(3-\mu)}\;,\eqno\stepeq$$ 
where $\alpha_{\rm M}$ is the fraction of the dissipated energy which goes into
magnetic fluctuations, $\zeta$ the fraction of the electrons which are
accelerated and $\mu$ the index of the fluctuation spectrum. 
With $1.5\le\mu\le2$, $\alpha_{\rm M}=0.1 - 1$, $\zeta\sim 10^{-3}$
and $\epsilon/m_{\rm e}c^2\sim 500$ (for $\Gamma_1/\Gamma_2=4$) values of
$\Gamma_{\rm e}$ in the range 
$10^3 - 10^4$ can be obtained.\par
A fraction of the synchrotron photons will be shifted to even higher energy
by inverse Compton scattering which now occurs in the limit where
$$w={\Gamma_{\rm e} E_{\rm syn}^0\over m_{\rm e}c^2}\simeq 33
\left({B\over 1000 {\rm G}}\right)\left({\Gamma_{\rm e}\over 10^4}\right)^3
\;,\eqno\stepeq$$ 
is large, $E_{\rm syn}^0=E_{\rm syn}/\Gamma_{\rm r}$ being the synchrotron energy
in the comoving frame. The inverse Compton photons have an energy
$$E_{\rm IC}^0\simeq\Gamma_{\rm e} m_{\rm e}c^2=5\left({\Gamma_{\rm e}\over
10^4}\right)\ \ {\rm GeV}\;,\eqno\stepeq$$ 
in the comoving frame and carry a fraction 
$$\alpha_{\rm IC}={\tau_{\star} \Gamma_{\rm e}^2/w\over 1+\tau_{\star} 
\Gamma_{\rm e}^2/w}\;,\eqno\stepeq$$ 
of the dissipated energy. The optical depth $\tau_{\star}$ has to be computed
with the Klein-Nishina cross-section which, in the limit $w\gg 1$, gives
$$\tau_{\star}\simeq \left[{\kappa_{\rm T} {\dot M}_{\rm shock} t_{\rm syn}
\over 4\pi r_{\star}^2 (1+\tau_{\star}\Gamma_{\rm e}^2/w)}\right]
\left({3\over 8 w}\right)
\left[1+{\rm ln}(2 w)\right]\;,\eqno\stepeq$$ 
and therefore
$$
\displaylines{
\hbox to \hsize{}\cr
{\tau_{\star}\Gamma_{\rm e}^2\over w}
\left(1+{\tau_{\star}\Gamma_{\rm e}^2\over w}\right)\simeq
8\;10^{-4}[1+{\rm ln}(2w)]
\left({{\dot E}\over 10^{52}\;{\rm erg.s}^{-1}}\right)\times\hfill\cr
\ \ \ \ \left({t_{\rm var}\over 1\;{\rm s}}\right)^{-2}
\left({{\bar \Gamma}\over 300}\right)^{-6}
\left({B\over 1000 {\rm G}}\right)^{-4}
\left({\Gamma_{\rm e}\over 10^4}\right)^{-5}\;.\hfill\stepeq\cr
}
$$
The very large exponents which appears in (29) show that a small variation
of the parameters can induce large changes in the relative importance of the 
synchrotron and inverse Compton processes. In practice, we find that $\alpha_{\rm IC}$
is gene\-ral\-ly small in the early part of a burst but increases at later times 
essentially due to the reduction of the equipartition magnetic field in
shocks at large distances from the source (see Sect. 3.1 below).\par
To be efficient the emission process must also occur on a time scale 
$t_{\rm em}$
shorter than the shell expansion time $t_{\rm ex}$ (Eq. 2). Using 
$r\sim c t_{\rm var} {\bar \Gamma}^2$, this condition becomes
$${t_{\rm syn}\over 1+ Q_{\rm IC}}<{\bar \Gamma} t_{\rm var}\;,\eqno\stepeq$$ 
where $Q_{\rm IC}=\tau_{\star}\Gamma_{\rm e}^2$ (resp. $\tau_{\star}
\Gamma_{\rm e}^2/w$) for $w\ll 1$ (resp. $w\gg 1$). With expressions
(11), (13), (14) and (20) above, (30) can be written
$$
\displaylines{
\hbox to \hsize{}\cr
2\;10^{-4}\alpha_{\rm B}^{-1} (\Gamma_{\rm int}-1)^{-1}
\left({{\dot E}\over 10^{52}\;{\rm erg.s}^{-1}}\right)^{-1}\times\hfill\cr
\ \ \ \ \left({t_{\rm var}\over 1\;{\rm s}}\right)
\left({{\bar \Gamma}\over 300}\right)^5
\left({\Gamma_{\rm e}\over 10^4}\right)^{-1}<1+Q_{\rm IC}\;.\hfill\stepeq\cr
}
$$
It is more easily satisfied when gamma-rays directly come from 
synchrotron emission since then $\Gamma_{\rm e}\sim 10^4$. However, even
for $\Gamma_{\rm e}\sim 10^2$ the emission time remains smaller than the
expansion time as long as the shocks
are sufficiently strong ($\Gamma_1/\Gamma_2\ga 2$).\par
\beginfigure{3}
\caption{{\bf Figure 3.} Physical parameters governing the emission mechanism:
(a) fraction $\alpha_{\rm syn}=1-\alpha_{\rm IC}$ of the energy which is 
radiated 
by the synchrotron process; (b) Lorentz factor $\Gamma_{\rm e}$ of the relativistic electrons;
(c) equipartition magnetic field $B_{\rm eq}$; (d) synchrotron energy 
$E_{\rm syn}$. In (a), (b) and (d) the thick lines correspond to $\Gamma_{\rm e}$
given by Eq. 24 and the thin lines to $\Gamma_{\rm e}=10^4$.}
\endfigure
Some results of a model where GRBs are produced by the inverse Compton process
have been already presented elsewhere (Mochkovitch \& Fuchs, 1996). In this study we limit
ourselves to synchrotron emission models except when we discuss the optical
properties which strongly differ between the two cases. 
In synchrotron emission models the total efficiency for the conversion of wind kinetic energy to
gamma-rays below a few MeV is given by 
$$f_{\rm{tot}}=f_{\rm d}\times \alpha_{\rm e}(1-\alpha_{\rm IC})\;.\eqno\stepeq$$ 
where $\alpha_{\rm e}$ is the fraction of the dissipated energy which is 
transferred to the electrons. According to Bykov \& M\'esz\'aros (1996) 
$\alpha_{\rm e}$ is comparable
to the fraction $\alpha_{\rm M}\sim 0.1 - 1$ 
of the energy which is initially
injected into magnetic fluctuations. The total efficiency therefore does not
exceed a few percent which imposes severe constraints on the energy source
or/and system geometry. \par
The energy available from disc accretion to a 
black hole
$$E\simeq {1\over 6} M_{\rm D} c^2 = 3\ 10^{53}\left({M_{\rm D}\over M_{\odot}}\right)
\ \ {\rm erg}\;,\eqno\stepeq$$ 
where $M_{\rm D}$ is the disc mass, can be less than $10^{53}$ erg for the
coalescence of two neutron stars since numerical simulations indicate 
that $M_{\rm D}\sim 0.2 - 0.3$ M$_\odot$ (Rasio \& Shapiro 1994;
Davies et al. 1994; Ruffert, Janka \& Schaefer 1996). The conversion of disc 
gravitational energy into wind kinetic energy should then be very efficient 
in order to 
account for cosmological GRBs. Naturally, if the wind is beamed 
in a solid angle $\delta\Omega$ along the system axis 
the energy requirement is smaller by a factor ${\delta\Omega\over 2\pi}$ but one 
has now to face a statistical problem since even optimistic estimates
of the neutron star merging rate are not considerably larger than the burst 
rate (Phinney 1991; Narayan, Piran \& Shemi 1991; Tutukov \& Yungelson 1993; 
Lipounov et al. 1995).
The situation is less critical if the disc results from the disruption of
a neutron star by a black hole or the collapse of a 
massive star because the mass of the disc can 
be $M_{\rm D}\ga 1$ M$_\odot$ (and even $M_{\rm D}\ga 10$ M$_\odot$).
\beginfigure{4}
\caption{{\bf Figure 4.} Burst profiles for the initial distribution of the Lorentz factor 
shown in Fig.1. The count rate $C_{2+3}$ (in arbitrary unit) is given in the interval
50 -- 300 keV, corresponding to BATSE bands 2 and 3; (a) profile obtained 
with Eq. 24 for the electron Lorentz factor; (b) same as (a) with 
$C_{2+3}$ in logarithmic scale which illustrates the exponential decay 
after maximum; (c) profile obtained with a constant 
$\Gamma_{\rm e}=10^4$; (d) same as (c) with $C_{2+3}$ in logarithmic scale.}
\endfigure

%
%

\section{Temporal properties}
\subsection{Burst profiles}
We first study the temporal properties of our burst models when the Lorentz
factor in the wind has the simple shape shown in Fig.1. 
We inject an energy ${\dot E}={10^{52}\over 4\pi}$ erg.s$^{-1}$.sr$^{-1}$ 
which, for an efficiency of conversion into gamma-rays of a few percent,
yields a total energy $E\simeq {10^{51}\over 4\pi}$ erg.sr$^{-1}$
 for a burst lasting a few seconds.
We obtain the 
burst profiles (in number of photons per second between 50 and 300 keV,
which corresponds to BATSE bands 2 and 3) by adding the contributions of
all the internal elementary shocks which occur during the expansion of the wind
as explained in Sect. 2. For each shock
we compute 
$\alpha_{\rm syn}=1-\alpha_{\rm IC}$ (Eq. 27) where $\alpha_{\rm IC}$
is the fraction of the energy which goes to inverse Compton photons,
the equipartition magnetic field 
$B_{\rm eq}$ (Eq. 13 with $\alpha_{\rm B}=1/3$),
the electron Lorentz
factor $\Gamma_{\rm e}$ (Eq. 24 with ${\alpha_{\rm M}\over \zeta}=1000$ and
$\mu=1.75$)
and the synchrotron 
energy $E_{\rm syn}$ (Eq. 15). 
These quantities are represented in Fig.3 as a function of arrival time
$t_{\rm a}$. As in Fig.2 there are two
branches corresponding to the forward and reverse shocks. Typical values
are $10^2 - 10^4$ G for the magnetic field, $2 - 20\times10^3$ for the electron Lorentz factor and 
10 keV -- 1 MeV for the synchrotron 
energy. 
\beginfigure{5}
\caption{{\bf Figure 5.} Full line: ratio of the decay time to the rise time of burst profiles
obtained with an initial distribution of the Lorentz factor homothetic to
that shown in Fig.1 but with different durations.
For $t_{90}<1.65$ s the profiles decay faster than they rise. Dashed line: 
the discontinuity in the initial distribution 
of the Lorentz factor has been replaced by a smoother
transition (Eq. 36). Now, only bursts with $t_{90}<0.65$ s decay faster than they rise.}
\endfigure
\beginfigure{6}
\caption{{\bf Figure 6.} Burst profiles for three initial distributions of the Lorentz factor
in the wind. In all three cases a rapid component with $\Gamma=400$ is 
decelerated by a series of slower layers. The masses in the rapid component 
and slower layers are comparable. (a) Relatively simple profile with four 
layers which produce four intensity pulses two of which partially overlap;
(b) and (c) more complex profiles with 15 slow layers. Notice that the distributions
of the Lorentz factor in (b) and (c) are homothetic, (c) being simply ten times longer
than (b).}
\endfigure
\beginfigure{7}
\caption{{\bf Figure 7.} Same as Fig.6 with a random fluctuation (of maximum amplitude
$\pm$ 20\%) added to the average value $\Gamma=400$ of the rapid component
of the wind. The resulting profiles now exhibit 
variability on a short time scale.}
\endfigure
The photons produced in the elementary shocks
are distributed according to a synchrotron spectrum
$${{\rm d}n(E)\over {\rm d}E}\propto {e\over E_{\rm syn}}\left({E\over
E_{\rm syn}}\right)^{-x}\;,\eqno\stepeq$$ 
with $x=2/3$ for $E<E_{\rm syn}$, $2<x<3$ for $E>E_{\rm syn}$ and where
$e$ is the energy which is dissipated in the shock (Eq. 5).
The resulting burst profile is shown in Fig.4a for a high e\-ner\-gy index
$x=2.5$. Cosmological
effects (time dilation and redshift) have been taken into account, assuming
that the burst is located at $z=0.5$. 
The duration $t_{90}$ which,
accor\-ding to the definition used for the BATSE data is the time
during which 90\% of the total fluence is received (excluding the first and
last 5\%) is 10.04 s, very similar to the duration of 
wind emission $t_{\rm w}$. 
The decay after maximum is close to an exponential as can be seen in the plot
of the logarithm of the count rate versus time in Fig.4b where 
there is a quasi linear decline after maximum between $t\simeq 12$ s
and $t\simeq 20$ s. However the rise time in our profile is not
much shorter than the decay time and the burst is therefore not very
dissymmetric if we except the low intensity exponential tail after $t= 15$ s. 
We found that a profile much closer to the characteristic FRED (Fast
Rise Exponential Decay) shape observed in many bursts (or in 
individual pulses inside a complex burst) can be obtained by assuming
that the electron Lorentz factor varies more slowly with $\epsilon$ (the 
dissipated energy per proton) than a power-law of index $1/(3-\mu)$ (Eq. 24).
This would be the case if the fraction $\zeta$ of accelerated
electrons increased with $\epsilon$. Such a behavior is observed in simulations
of collisionless non relativistic shocks (Bykov, private 
communication) and we have supposed that it remains valid in the relativistic
limit. Moreover, we have made the simple choice $\zeta\propto\epsilon$
which leads to a characteristic Lorentz factor for the electrons which is 
independent of $\epsilon$. The synchrotron energy in the elementary 
shocks has been represented in Fig.3d for a constant $\Gamma_{\rm e}=10^4$ and the
corresponding burst profile is shown in Fig.4c. It has now a typical
FRED shape 
with a ratio of the decay time to the rise time 
${\tau_{\rm d}\over\tau_{\rm r}}=3.4$ where $\tau_{\rm r}$ and $\tau_{\rm d}$
are defined respectively by
$$\tau_{\rm r}=t_{\rm max}-t_{5\%}\ \ ,\ \ \tau_{\rm d}=t_{95\%}-t_{\rm max}\;,
\eqno\stepeq$$ 
where $t_{\rm max}$ is the time of maximum count rate and $t_{5\%}$ (resp. 
$t_{95\%}$) the time when 5\% (resp. 95\%) of the total fluence has been 
received.
We have tested with our model the tendency for short bursts (or short
pulses within a complex burst) to become 
more symmetric (Norris et al. 1996). We have computed the profiles obtained
when the duration of wind emission $t_{\rm w}$ is varied while the initial 
distribution of the Lorentz
factor remains homothetic to a given shape for which we choose the one 
represented in Fig.1 where the wind consists of a slow part ($\Gamma=100$) 
followed
by a rapid part ($\Gamma=400$) both containing the same mass.
The results (for $z=0.5$) are shown in Fig.5 where we have plotted
the ratio ${\tau_{\rm d}\over\tau_{\rm r}}$ 
as a function of $t_{90}$. It appears that ${\tau_{\rm d}\over\tau_{\rm r}}$ 
decreases from about 3 when $t_{90}\simeq10$ s to about 0.3 when
$t_{90}\la 0.5$ s with ${\tau_{\rm d}\over\tau_{\rm r}}= 1$ for
$t_{90}\simeq 1.65$ s. 
We therefore reproduce the observed behavior but the effect is even 
exagerated
for the shortest pulses which decay faster than they rise. 
We believe that this might be a consequence of the crude assumptions made in
our simple model and we expect that more detailed hydrodynamical simulations
(Daigne and Mochkovitch, in preparation) could help to improve the profiles.
Already, we found that better results can be obtained (with
${\tau_{\rm d}\over\tau_{\rm r}}= 1$ for $t_{90}\simeq 0.65$ s) 
if the discontinuity
between the two extreme values of $\Gamma$ is replaced 
by a smoother transition of the form
$$\Gamma(t/t_{\rm w})=250+150\;{\rm cos}[2.5\pi(t/t_{\rm w}-0.6)]\;,\eqno\stepeq$$ 
for $0.6\le t/t_{\rm w}\le 1$ and $\Gamma=400$ for $t/t_{\rm w}<0.6$.
However, the efficiency $f_{\rm d}$ of the dissipation process is then reduced
by nearly a factor of 2.\par
Norris et al. (1996) have shown that in most cases complex bursts can be analyzed
in terms of a series of (possibly overlapping) pulses. 
In the same way we build complex
bursts with our model by the addition of intensity pulses formed in the deceleration of
rapid parts of the wind by slower ones which were emitted previously. We suppose that
the initial distribution of the Lorentz factor is made of a rapid component
(with an average value of $\Gamma$ of a few hundreds) and of some slower
layers (with $\Gamma\simeq 100$). The total mass in the slow layers has to be
comparable to the mass in the rapid component in order to keep the efficiency 
at a reasonable level. A few examples of synthetic profiles are presented in 
Fig.6. It can be seen that a great diversity of burst shapes 
can be obtained if the distribution of the Lorentz factor in the wind 
varies from one event to the other.
\beginfigure{8}
\caption{{\bf Figure 8.} Spectrum of the burst corresponding to the profile of Fig.7a.
The number of photons per energy interval $n(E)$ and the product $E^2 n(E)$
are shown in arbitrary units. The dashed line is a fit of the spectrum with 
Band's formula
in the interval 10 keV -- 10 MeV. The product $E^2 n(E)$ is maximum at the
peak energy $E_{\rm p}=365$ keV.}
\endfigure

\subsection{Short timescale variability}
The profiles shown in Fig.4 and 6 have a satisfactory ge\-ne\-ral shape but do not
exhibit any variability on a short time scale. Rapid 
temporal variations cannot result from small irregularities of the emitting
surface since photons coming from many different regions in space and time are received
at a same time by the observer which leads to a loss of co\-he\-ren\-ce of the 
temporal variations (Woods \& Loeb 1995; Sari \& Piran 1997b,c). Instead, one has to consider again that the Lorentz
factor itself can fluctuate at the millisecond level. This is not unrealistic 
if the flow at the origin of the wind is very irregular and turbulent
since one millisecond cor\-res\-ponds to the typical dynamical time scale of a disc
orbiting a stellar mass black hole. Nothing being known about 
the temporal spectrum of the fluctuations
we have simply added a random fluctuation to the Lorentz factor
of each of the layers initially injected (every 2 ms) in the wind. 
The adopted amplitude for these fluctuations is 10 -- 20\% of the average value of 
$\Gamma$. The resulting profiles represented in Fig.7 now show the 
rapid temporal 
variations which are seen in most observed bursts.

%
%

\section{Spectral properties}
\subsection{Burst spectrum}
The overall burst spectrum is the sum of all the elementary contributions 
(Eq. 34)
from the internal shocks. The spectrum corresponding to the profile shown in
Fig.7a
is represented in Fig.8 (again a cosmological redshift $z=0.5$ has been assumed).
Its shape can be easily understood: let $E_{\rm syn}^{\rm min}$ and
$E_{\rm syn}^{\rm max}$ be the minimum and maximum of the synchrotron energy
for the whole set of elementary shocks. 
As long as $E$ is smaller (resp. larger) than
$E_{\rm syn}^{\rm min}$ (resp. $E_{\rm syn}^{\rm max}$) 
the number
of photons $n(E)$ per unit energy interval is a power-law 
of index $-2/3$ (resp. $-2.5$) and in the intermediate region the 
current index evolves from $-2/3$ to $-2.5$ as $E$ goes beyond the value of
$E_{\rm syn}$ for a growing number of elementary shocks.
In the four BATSE bands, between 20 keV and a few MeV, the spectrum 
can be well described with Band's formula (Band et al. 1993)
$$
\displaylines{
\hbox to \hsize{}\cr
n(E)= A\left({E\over 100\;{\rm keV}}\right)^{\alpha}{\rm exp}\left(-{E\over
E_0}\right)\ \ \ {\rm for}\ \ \ (\alpha-\beta)E_0\ge E\;,\hfill\cr
n(E)= A\left[{(\alpha-\beta)E_0\over 100\;{\rm keV}}\right]^{\alpha-\beta}
{\rm exp}(\beta-\alpha)\left({E\over 100\;{\rm keV}}\right)^{\beta}\hfill\cr
{\rm for}\ \ \ (\alpha-\beta)E_0\le E\;.\hfill\stepeq\cr
}
$$
The parameters $\alpha$, $\beta$ and $E_0$ 
have been adjusted to obtain the best possible fit of the spectrum in Fig.8.
We get $\alpha=-1.33$, $\beta=-2.31$ and $E_0=544$ keV, in agreement with
typical
values found in observed bursts.
The product $E^2 n(E)$ is also shown in 
Fig.8. It is maximum at the peak energy $E_{\rm p}= 365$ keV where the bulk 
of the emission takes place.
\beginfigure{9}
\caption{{\bf Figure 9.} Duration-hardness ratio (HR$_{32}$) relations for a simple (one pulse) burst and
a complex burst with five pulses. The relations are shown for three values of the
high energy index $x=2$, 2.5 and 3 of the elementary synchrotron spectrum
and two redshifts $z=0.3$ and 1. It can be seen that the effect of the 
redshift is negligible.}
\vskip -0.2cm
\endfigure

\subsection{Duration-hardness ratio relation}
Shorter bursts are expected to be harder in our model. Internal shocks
are formed at an approximate radius $r\sim c t_{\rm var}{\bar \Gamma}^2$
and are therefore closer to the source if the burst evolves on a short
time scale. Assuming that the
injected power ${\dot E}$ is independent of the duration of wind injection 
$t_{\rm w}$ 
the equipartition magnetic field is stronger and the synchrotron energy is
larger in the dissipation region for shorter bursts. To obtain the 
duration-hardness relation we compute the hardness ratio 
${\rm HR}_{32}$ (defined as the 
ratio of the number of photons received in BATSE band 3 to that in band 2)
for bursts with homothetic initial distributions of the Lorentz factor
but different values of $t_{\rm w}$. For a given duration $t_{90}$
the hardness ratio is a function 
of the high energy index of the elementary spectrum (Eq. 34)
(the bursts becoming softer when $x$ increases from 2 to 3),
of the detailed history of the Lorentz factor
and of the cosmological redshift $z$. 
Complex bursts tend to be harder because the
distribution of the Lorentz factor varies on a shorter time scale 
at given $t_{\rm w}$
and dissipation therefore begins earlier than in more regular bursts of same 
duration. Finally cosmological effects shift bursts in
the ${\rm HR}_{32}$ -- $t_{90}$ diagram downward (redshift) and to the right (time
dilation). 
However the shift is nearly colinear to the duration-hardness relation 
which therefore remains practically unchanged at different $z$.
This is illustrated in Fig.9 where the
duration-hardness relation has been represented for several
values of $x$ and $z$ and for a simple (single pulse) and a complex
burst. 
In agreement with the observations (Kouveliotou et al. 1993; 
Dezalay et al. 1996) a transition occurs at $t_{90}\sim$ 2 s.
The shortest bursts reach a limit 
$${\rm HR}_{32}\simeq {300^{1/3}-100^{1/3}\over 100^{1/3}-50^{1/3}}\simeq 2.1\;,\eqno\stepeq$$ 
when $E_{\rm p}$ is large enough for both BATSE bands 2 and 3 to lie in
the region where $n(E)\propto E^{-2/3}$.
The longer bursts tend to various limiting values of the hardness ratio depending 
on the choice made for the high energy index $x$. 
\beginfigure{10}
\vskip -0.5cm
\caption{{\bf Figure 10.} Evolution of hardness with time: the instantaneous value of the 
peak energy $E_{\rm p}$ (thick line) is represented together with
the burst profile (thin line). Upper panel: for the simple one pulse
burst of Fig.4c. Lower panel: for the more complex burst of Fig.6a.}
\vskip -0.4cm
\endfigure

\subsection{Spectral evolution}
\subsubsection{Instantaneous hardness}
The spectral evolution of GRBs shows a few trends which are followed by a 
majority of bursts but also suffer some exceptions and are therefore
not universal (Bhat et al. 1994; Ford et al. 1995). First, spectral hardness and count rate appear
to be correlated. Within intensity pulses both increase and decrease 
together, the hardness usually preceding the count rate. Another
trend is a global hard-to-soft evolution over the course of the burst
outside intensity pulses. Finally, later pulses tend to be softer than
earlier pulses, even if they have a greater intensity.
\par
We have compared the spectral evolution of our synthetic burst models to
these observational results.
The hardness can be obtained as a function of time through the estimation of 
the instantaneous value of $E_{\rm p}$, the energy of the peak of $E^2n(E)$.
We first considered the simple burst of Fig.4c which has a characteristic
FRED shape and a duration $t_{90}=10.23$ s. The hardness and count rate 
evolve similarly (see Fig.10) but their maxima are separated by a time lag of 0.89 s, the maximum of
$E_{\rm p}$ occurring before that of the count rate. In complex 
bursts the hardness increases during intensity pulses and also precedes the
count rate (Fig.10).\par
\beginfigure{11}
\vskip -0.5cm
\caption{{\bf Figure 11.} Upper panel: normalized profiles for the burst of
Fig.4c now 
represented in all four BATSE bands. Lower panel: half maximum widths of 
the profiles as a function of energy. The fit of the model results (dashed line) has a slope
$p=-0.39$.}
\vskip +0.5cm
\endfigure
The correlation between spectral hardness and count rate is
therefore correctly reproduced in our models. We encountered more difficulties
with the global hard-to-soft evolution which 
was observed in 70\% of the sample of bright long bursts studied by
Ford et al. (1995). In synthetic bursts (Fig.10) it is present as long as 
the profiles remain relatively
simple (i.e. dominated by one main pulse or made of just a few pulses) 
while in complex bursts
with many pulses only the correlation between hardness and count rate 
is clearly visible. 
Also, the hardness of successive pulses remains essentially 
correlated to the intensity
instead of decreasing like in 50\%
of the Ford et al. (1995) sample.

\subsubsection{Pulse shape as a function of energy}
When observed in spectral bands of increasing energy, pulses become narrower as shown by
Norris et al. (1996) who a\-na\-lyzed a large number of pulses in the four 
BATSE bands. They found that their (half maximum) width can be well represented 
by a power-law
$$W(E)\propto E^{-0.4}\;.\eqno\stepeq$$ 
The same relation was obtained by Fenimore et al. (1995) who used the 
autocorrelation of averaged
burst profiles 
instead of individual pulses. 
In our synthetic burst models the pulse width also decreases at high energy. 
Figure 11 represents a single pulse burst in the four BATSE bands. The
width can be fitted by power-laws such (39) but
not with a unique exponent $p= -0.4$ for all the pulses. We indeed
get $p\simeq -0.4$ for pulses of 2 -- 10 s but for pulses of 
0.1 -- 1 s, $p\ga -0.2$. Norris et al. (1996) 
obtained $p = -0.4$ as an average on a collection of pulses with
duration ranging from 0.1 to 10 s but do not provide the value of $p$ for the
shorter and longer pulses separately which does not allow a detailed
comparison with our theoretical results.

\subsection{The $E_{\rm p}$ - fluence relation}
Liang \& Kargatis (1996) discovered in a sample of 37 BATSE bursts
an exponential dependence of the peak
energy $E_{\rm p}$ on photon fluence $F_{\rm ph}$ during the declining part of
intensity pulses i.e.
$$E_{\rm p}\propto {\rm exp}(- aF_{\rm ph})\;.\eqno\stepeq$$ 
The photon fluence is defined as the integral of the photon flux 
from the beginning of the burst and $a$ is 
the slope of the ${\rm Log} E_{\rm p}$ -- $F_{\rm ph}$ relation. 
In complex bursts the slope stays approximately constant from 
one pulse to another. Synthetic bursts also follow a relation such (40)
but the slope in successive pulses
can somewhat vary, especially if they have very different 
intensity or duration (see Fig.12).
\beginfigure{12}
\vskip -0.3cm
\caption{{\bf Figure 12.} $E_{\rm p}$-fluence relation (thick line) 
for the burst of Fig.6a. The peak energy is represented in logarithmic scale
to show the section of 
linear decline following the moment of maximum count rate in 
the four intensity pulses.
The photon fluence corresponds to the integrated photon flux in BATSE bands 
2 and 3. The thin line shows the count rate in the same bands.}
\endfigure

\subsection{X-ray and optical counterparts}
The X-ray and optical counterparts recently discovered in two GRBs 
have been interpreted
in the context of cosmological models as the emission coming from a
relativistic shell expanding in the interstellar medium 
(Wijers, Rees \& M\'esz\'aros 1997; Vietri 1997; Waxman, 1997a,b).
If however GRBs are produced by internal shocks in a wind,
X-ray to optical photons should also be emitted together 
with the gamma-rays before the afterglow resulting from
the interaction with the interstellar medium. 
\par
To compute these early counterparts we consider a ty\-pi\-cal burst where an energy 
$E={2\;10^{52}\over 4\pi}$ erg.sr$^{-1}$ has been injected into the wind 
with a 5\% efficiency for the conversion to gamma-rays between 50 keV and
300 keV. For a GRB located at 2 Gpc ($z\sim 0.5$) the observed fluence in 
BATSE bands 2 and 3 will be $F_{\gamma}\simeq 2\;10^{-6}$ erg.cm$^{-2}$. This value
of $F_{\gamma}$ 
is then used to normalize a synthetic spectrum from which the
expected flux in the X-rays and visible can be finally obtained. 
In practice these fluxes are highly variable (as the gamma-rays) and have
been averaged over the duration $t_{90}$ of the burst. \par
We did not consider in this paper inverse Compton emission models but 
we now briefly discuss their optical properties since 
they greatly differ from synchrotron emission models. When the gamma-rays 
come from inverse Compton scattering, a fraction 
$\alpha_{\rm syn}\sim 1-\alpha_{\rm IC}$ of the total power is 
emitted at a typical synchrotron energy $E_{\rm p}/\Gamma_{\rm e}^2\sim 10 - 100$ eV where
$E_{\rm p}$ is the peak energy of the gamma-ray spectrum. 
The fraction $\alpha_{\rm syn}$ is fixed by the ratio $\alpha_{\rm e}/\alpha_{\rm B}$
through Eq. (17) and (23). Preliminary results 
indicate that for $\alpha_{\rm e}/\alpha_{\rm B}=1$
the emission in the visible of a burst of fluence $F_{\gamma}$ and duration
$t_{90}=10$ s could be as bright 
as $V=5 - 6$ which is already excluded by the limit set by the ETC and 
GROCSE instruments
(Krimm, Vanderspek \& Ricker 1996; Lee et al. 1997). With $\alpha_{\rm e}/\alpha_{\rm B}=100$ the predicted magnitude 
becomes $V=8 - 9$ still within the reach of ETC and GROCSE.
\par
In synchrotron emission models the optical counterpart is much weaker. Taking for example
the spectrum represented in Fig.8 which corresponds to a burst with $t_{90}=10.55\; \rm s$
we get a V magnitude of 18.4 for 
$F_{\gamma}\simeq  2\;10^{-6}$ erg.cm$^{-2}$.
A larger fluence would naturally produce
a brighter counterpart, possibly up to $V\sim 15 - 16$ for 
$F_{\gamma}\ga 2\;10^{-5}$ erg.cm$^{-2}$.
An optical emission in this luminosity range should be detectable 
by the next generation of counterpart search instruments, such as the 
TAROT project (Boer 1997). \par
In X-rays the calculated fluxes for the BeppoSAX Wide Field Camera 
are in reasonable agreement with the observations. In the case of
GRB 970228 
the X and gamma-ray fluences during the first 100 s are 
$F_{\rm X}(2 - 10\;{\rm keV})\simeq 1.2\;10^{-6}$ erg.cm$^{-2}$ and
$F_{\gamma}(40 - 700\;{\rm keV})\simeq 1.1\;10^{-5}$ erg.cm$^{-2}$ (Costa et al.
1997a).
For the same gamma-ray fluence 
the model predicts $F_{\rm X}^{\rm model}\simeq 0.7 - 1.1\;10^{-6}$ 
erg.cm$^{-2}$ depending on burst hardness
(the upper and lower limits of $F_{\rm X}^{\rm model}$ respectively
correspond to bursts with $E_{\rm p}=100$ and 350 keV).
\par
In GRB 970508 the peak X-ray flux and the gamma-ray fluence are 
${\cal F}_{\rm X}(2 - 10\;{\rm keV})\simeq 1.2\;10^{-8}$ erg.cm$^{-2}$.s$^{-1}$ 
and $F_{\gamma}(50 - 300\;{\rm keV})\simeq 1.1\;10^{-6}$ erg.cm$^{-2}$
(Costa et al. 1997b; Kouveliotou et al. 1997). 
The calculated X-ray flux for a burst of comparable hardness, 
${\cal F}_{\rm X}^{\rm model}\simeq 8\;10^{-9}$ erg.cm$^{-2}$.s$^{-1}$, is 
close to the
observed value while the initial X-ray emission predicted by afterglow
models appears to be an order of magnitude weaker (Sahu et al. 1997b).
This can be considered as an indication that both the gamma-rays and the 
initial X-ray emission are produced by internal shocks.
\par
The evolution of the afterglow which follows the emission 
from internal shocks can be obtained from a solution of the relativistic 
Sedov problem (M\'esz\'aros \& Rees 1997b).
Such a solution applies when the expanding shell of mass $M$ has swept 
up a mass $M_{\rm ISM}\sim {M\over\Gamma}$ in the interstellar medium 
which occurs after a deceleration time
$$t_{\rm dec}\simeq 180\;E_{52}^{1/3}n_1^{-1/3}\Gamma_2^{-8/3}\;{\rm s}\;,
\eqno\stepeq$$ 
where $E_{52}$ is the shell energy in units of ${10^{52}\over 4\pi}$ 
erg.sr$^{-1}$, $n_1$ the density of the interstellar medium in atom.cm$^{-3}$
and $\Gamma_2={\Gamma\over 100}$. Most of the energy
is radiated at the synchrotron frequency of the relativistic electrons
$$\nu_{\rm s}\simeq 3\;10^{16} \alpha_{\rm e}^2 \alpha_{\rm B} E_{52}^{1/2}
t_{\rm day}^{-3/2}\ \ {\rm Hz}\;,\eqno\stepeq$$ 
where $\alpha_{\rm e}$ and $\alpha_{\rm B}$ are the fractions of the 
dissipated energy which go to the electrons and magnetic field respectively. 
The flux at the synchrotron frequency is given by 
(Waxman, 1997a,b).
$${\cal F}_{\rm s}\simeq 6.5\;10^{-26}\sqrt{n_1}\alpha_{\rm B} E_{52} D_{\rm Gpc}^{-2}
\;\  {\rm erg.cm}^{-2}.{\rm s}^{-1}.{\rm Hz}^{-1}\;,\eqno\stepeq$$ 
and for $\nu > \nu_{\rm s}$ by
$$ {\cal F}_{\nu}={\cal F}_{\nu_{\rm s}}\left({\nu\over \nu_{\rm s}}\right)^{-\beta}\;,
\eqno\stepeq$$ 
where $\beta={{q-1}\over 2}$, $q$ being the exponent of the power-law
distribution of the accelerated electrons ($N(\Gamma_{\rm e})\propto
\Gamma_{\rm e}^{-q}$). The flux in the visible decreases from a maximum 
${\cal F}_{\rm V}^{\rm max}\simeq {\cal F}_{\rm s}$ following a power-law
of index ${-3\beta\over 2}$ the value of $\beta$ ($\sim  0.6 - 0.8$) being obtained
from  the observations.
For GRB 970508 a fit of the data also provides 
${\cal F}_{\rm V}^{\rm max}\simeq 6\;10^{-28}$ erg.cm$^{-2}$.s$^{-1}$.Hz$^{-1}$
(Sahu et al. 1997b).
With the gamma-ray fluence measured for this burst and assuming a
5\% efficiency for the conversion of wind kinetic energy to gamma-rays 
we obtain
$$\sqrt{n_1}\alpha_{\rm B}\simeq  3\;10^{-2}\;.\eqno\stepeq$$ 
If we adopt this value as typical we find that the afterglow 
of the burst (given in example above) with $E_{52}=2$ and $D_{\rm Gpc}=2$ 
has a V magnitude of 18.9 before the phase of power-law decline.
In inverse Compton emission models the initial optical counterpart
produced by internal shocks is considerably brighter than the afterglow
but in synchrotron emission models the
two contributions have comparable brightness. However, 
the optical signal will present a short interruption 
if the duration of the burst is smaller than the
deceleration time.
Conversely, 
if $t_{\rm w}\ga t_{\rm dec}$ the two contributions
overlap and form a continuous signal.
In any case, a detailed photometric follow-up in the optical range
beginning before the end of the gamma-ray burst would certainly provide crucial informations
about the emission mechanisms.

%
%

\section{Conclusions}
We have developed a simple model to compute the temporal and spectral properties 
of GRBs under the assumption that they originate from internal shocks in a 
relativistic wind. We have not discussed the critical point of how such a 
wind could form but the recent observations of optical counterparts for GRB 970228 and
GRB 970508 seem to indicate that a relativistic shell was indeed present 
in these objects.
\par
The distribution of the Lorentz factor in the wind has no reason to be uniform
and variations on several timescales (down to one millisecond which corresponds
to the dyna\-mi\-cal timescale of a relativistic disc orbiting a stellar mass 
black hole) can be expected. Layers of different velocities will collide
and form internal shocks within the relativistic wind and the energy dissipated
in these shocks can be emitted in the form of gamma-rays. We have followed the 
evolution of the relativistic wind using an approach where all pressure waves
are suppressed so that layers only interact through direct shocks. 
We assume
that the magnetic field reach equipartition in 
these shocks and that the electron Lorentz factor remains close to a constant
value $\Gamma_{\rm e}\sim 10^4$. This could be the case if, according to Bykov
\& M\'esz\'aros (1996) only a small fraction $\zeta$ of the electrons is 
accelerated in the shocks
and if $\zeta$ is also approximately proportional to the dissipated energy.
Gamma-rays are then directly produced by synchrotron emission
from the relativistic electrons.\par
This procedure allows us to construct synthetic bursts whose temporal and spectral properties
are compared to the observations. We obtain a series of encouraging results:
\par
1) It is possible to generate a great diversity of profiles simply by playing 
with the initial distribution of the Lorentz factor in the wind.\par
2) The profile of individual pulses is asymmetric, close to a ``FRED'' shape.\par
3) The short time scale variability of the profiles can be explained if the 
Lorentz factor itself varies at the millisecond level. 
\par
4) Synthetic spectra can be fitted with Band's formula with parameters 
$\alpha$, $\beta$ and $E_0$ comparable to those of observed bursts.
\par
5) The duration-hardness relation is a natural consequence of the model
since in short bursts dissipation occurs closer to the source where the
magnetic field and the synchrotron energy are larger.\par
6) Spectral hardness and count rate are correlated du\-ring burst evolution,
the hardness generally preceding the count rate.\par
7) The pulse width decreases with increasing energy following a power law
$W(E)\propto E^{-p}$ with $p\sim 0.4$ for pulses of duration 2 -- 10 s.\par
8) In the declining part of intensity pulses the peak energy $E_{\rm p}$ decreases 
exponentially with photon fluence.
\par
Some other properties of observed GRBs are however not so well reproduced 
by our model:\par
1) The shortest pulses tend to decay faster than they rise instead of being
symmetric.\par
2) Synthetic bursts do not show a global hard to soft evolution as frequently
as real bursts.
\par
3) The slope of the $E_{\rm p}$ -- fluence relation can differ among pulses 
inside the same burst.\par
It is not yet clear whether these difficulties represent real problems for
our model or are simply a consequence of some of the crude assumptions we have
made. In parti\-cu\-lar, the evolution of the wind should be followed with 
a detailed hydrodynamical code (Daigne and Mochkovitch, in preparation)
rather than with the simple method used here.\par
A more fundamental issue may be the rather low efficiency for the 
conversion of wind kinetic energy to gamma-rays. 
If GRBs result from the coalescence of neutron stars 
the wind cannot be strongly beamed since the merging rate 
is not very much greater than the burst rate. A large fraction of the 
energy released in the coalescence should therefore be injected into the wind.
\par
We believe that the present work has shown that if a relativistic wind carrying
enough energy can be produced, the internal shock model appears as a 
convincing candidate to explain GRBs.
Ways to both reach a high efficiency in the generation of the wind 
and to avoid at the same time baryonic pollution are among the difficult problems which
remain to be solved.

%
%

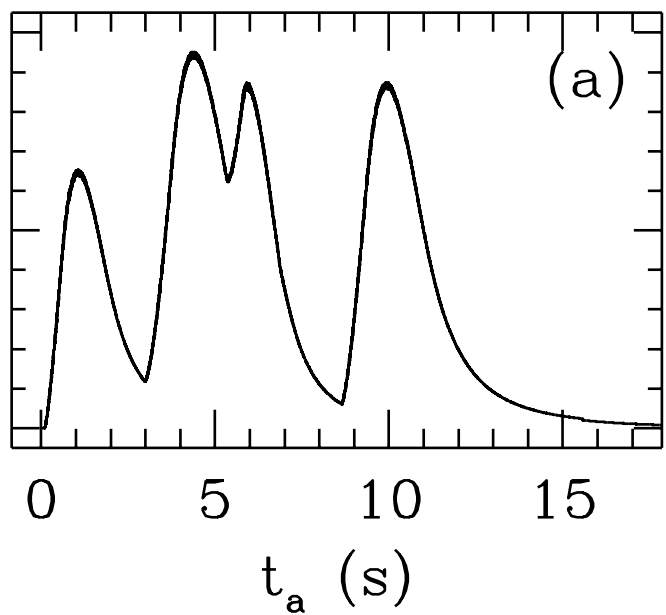
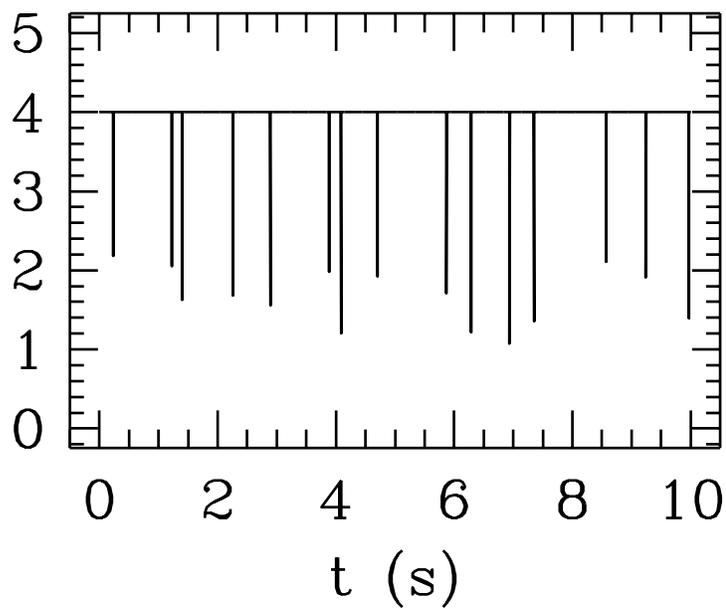
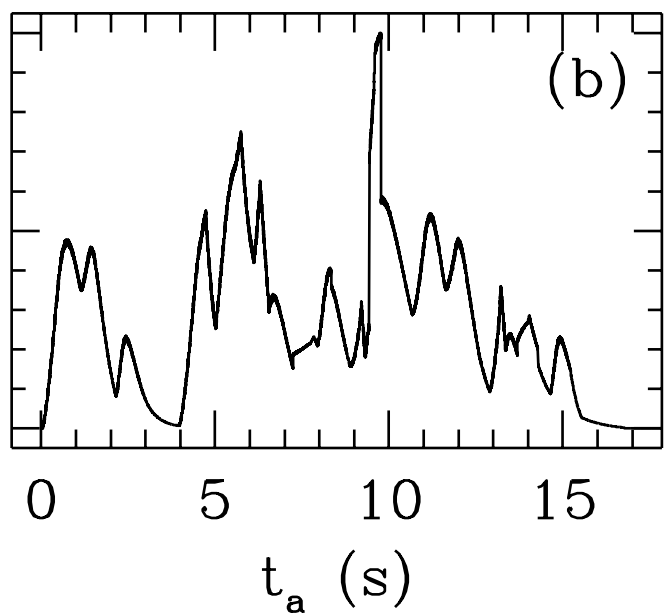
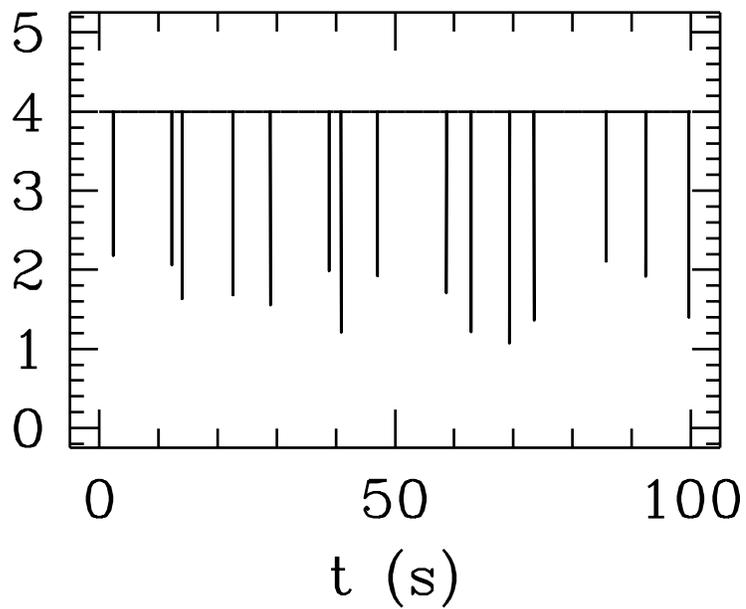
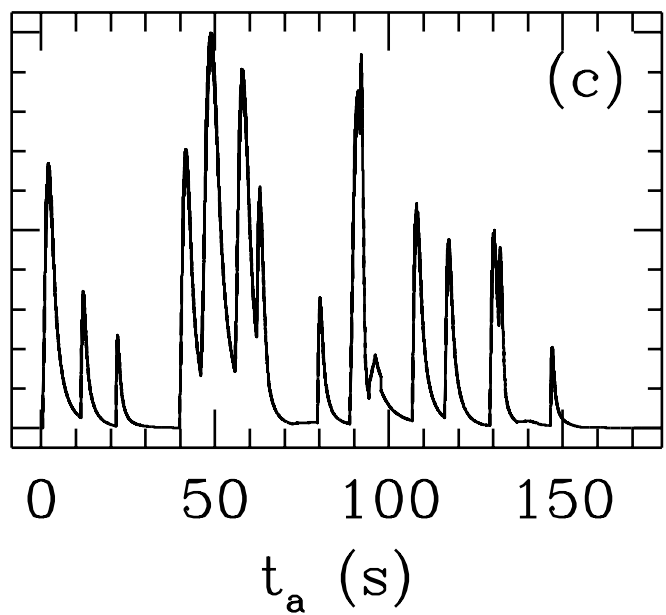
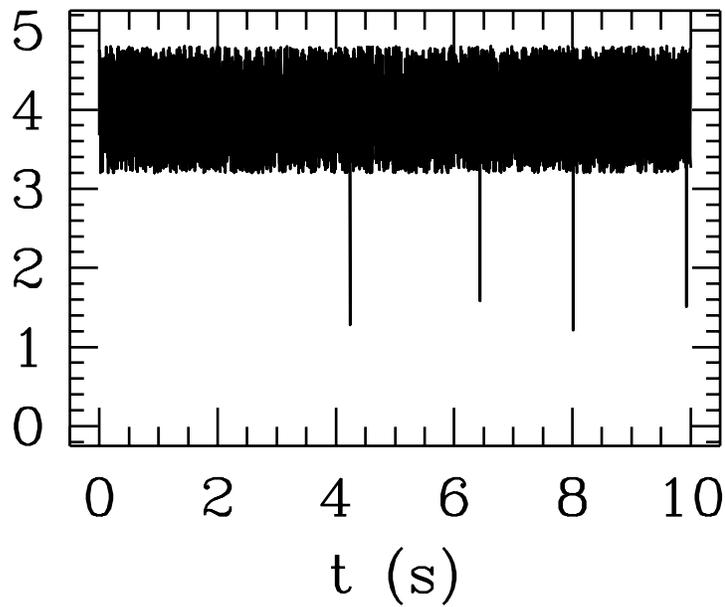
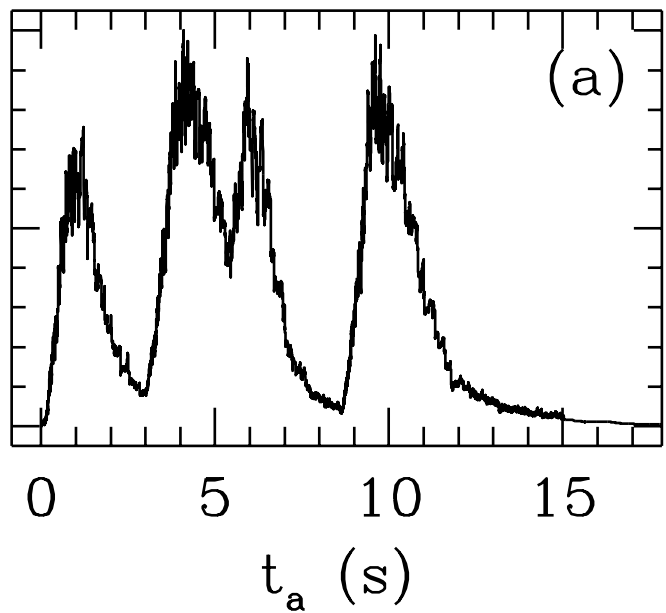
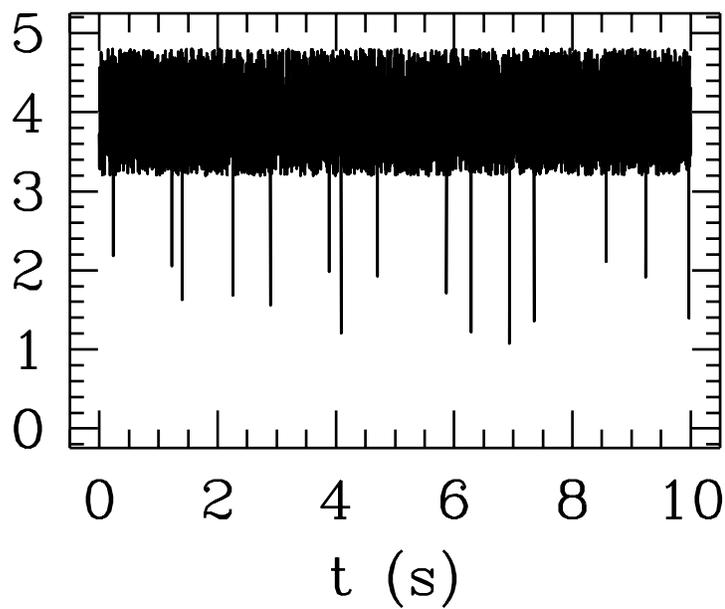
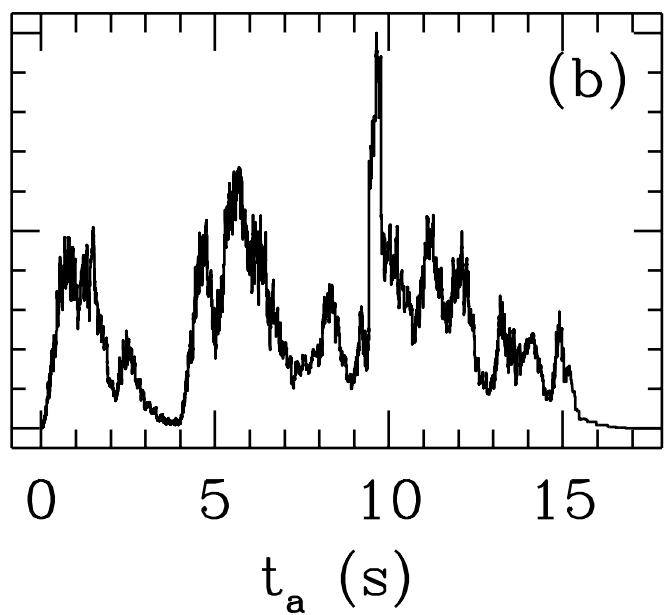
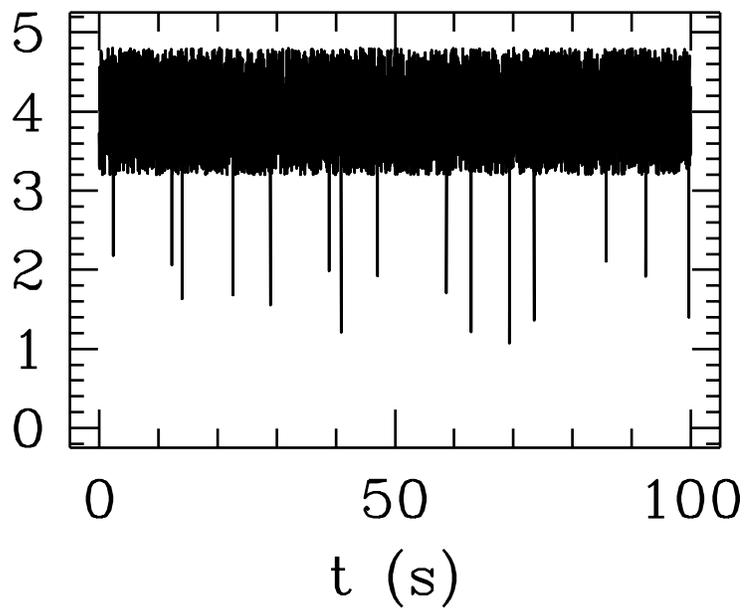
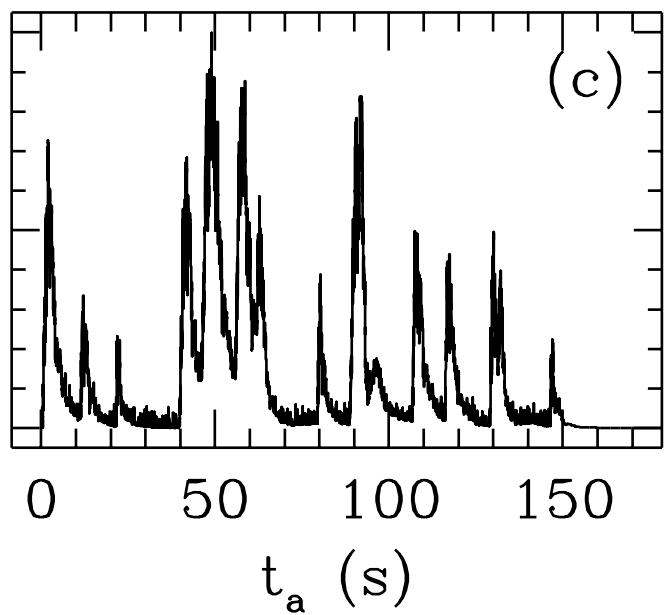
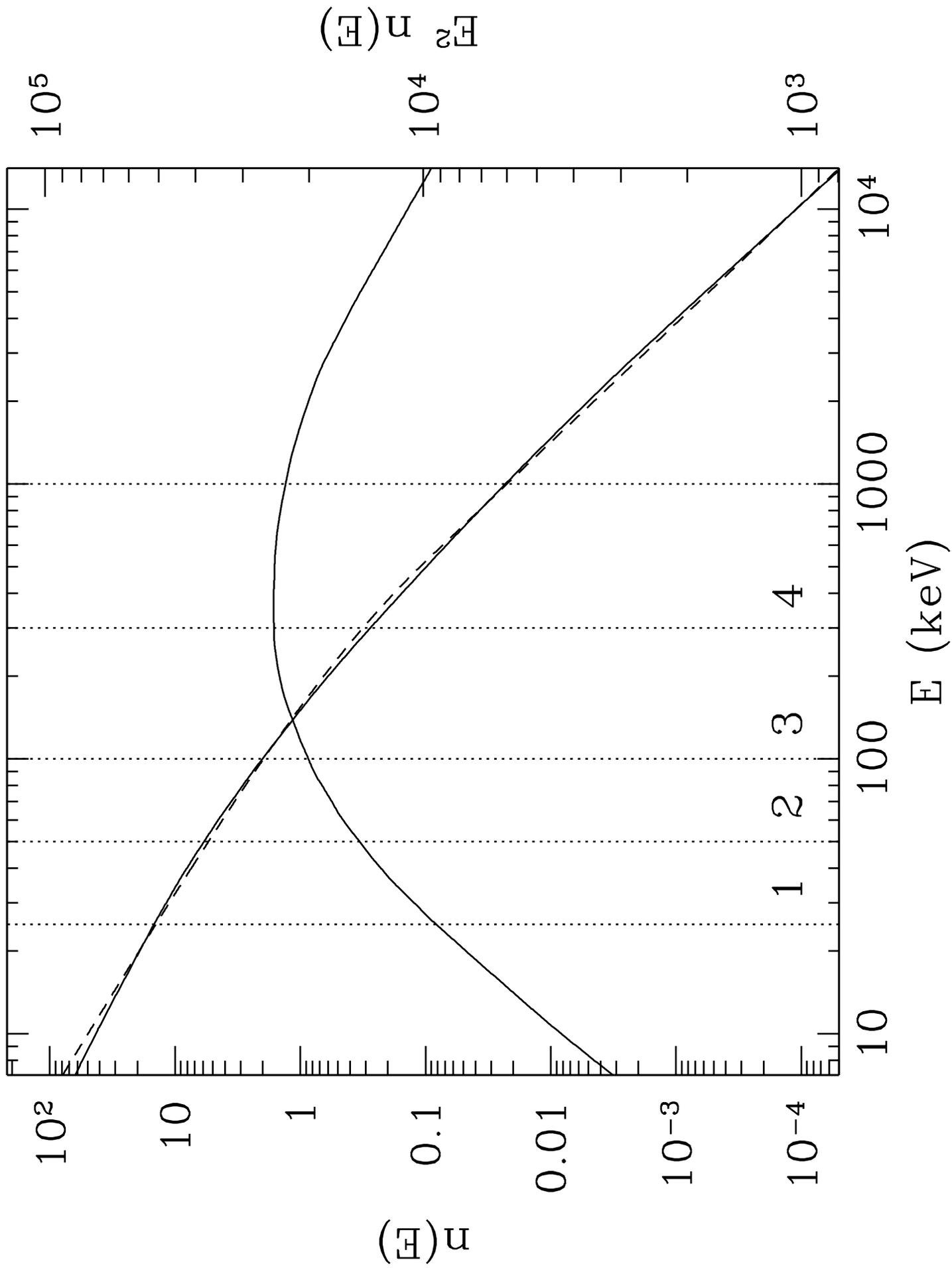
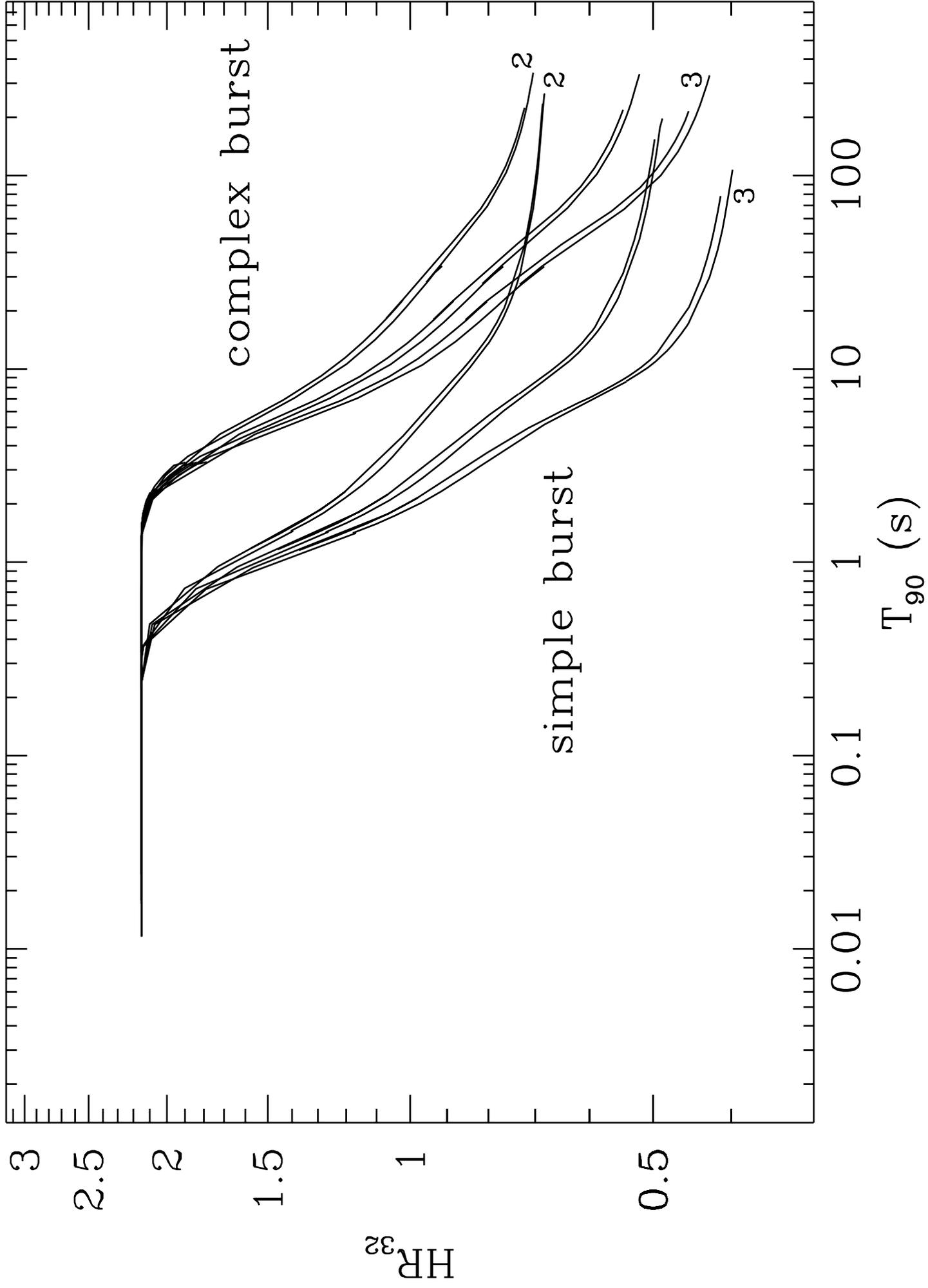
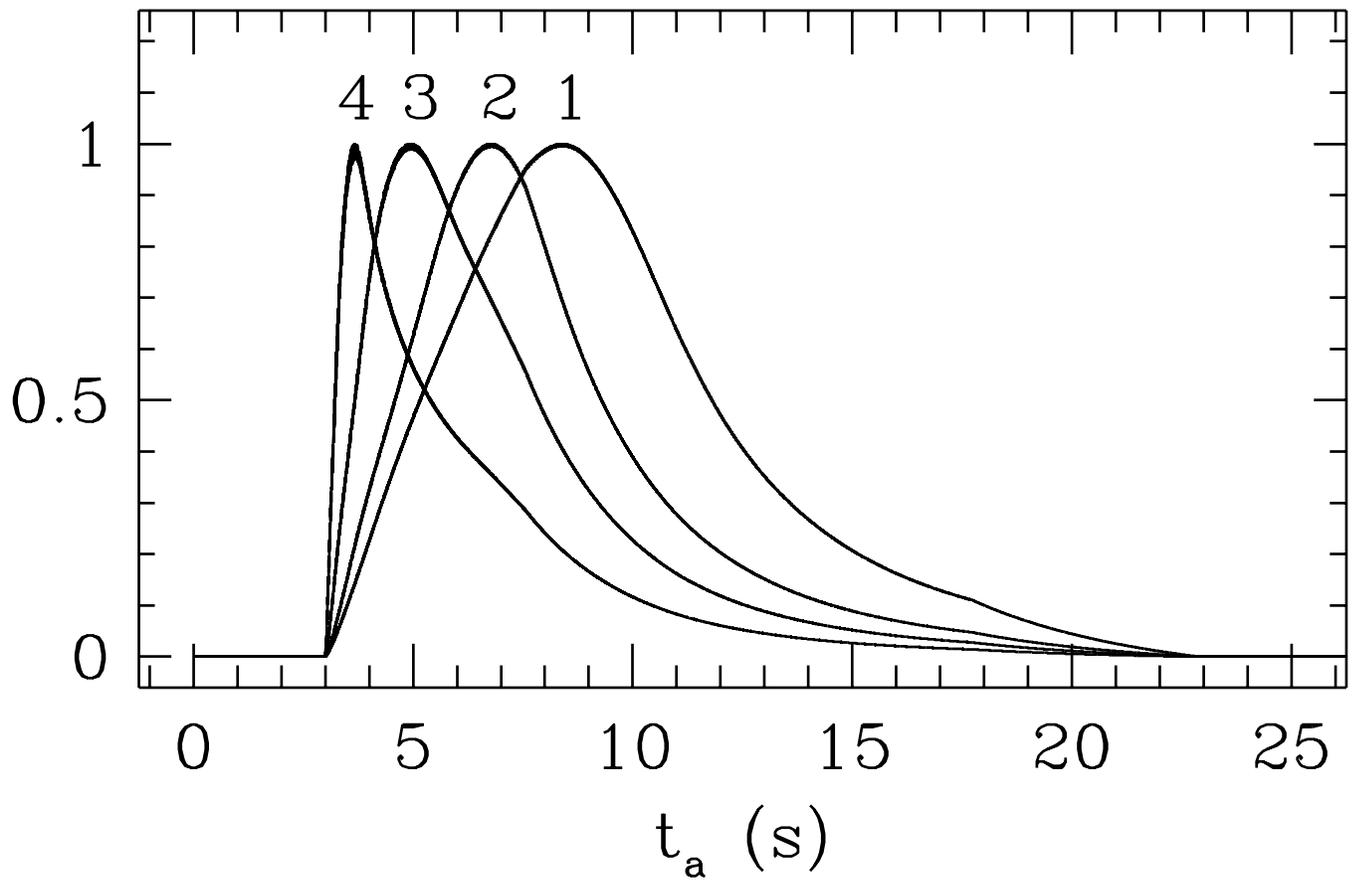
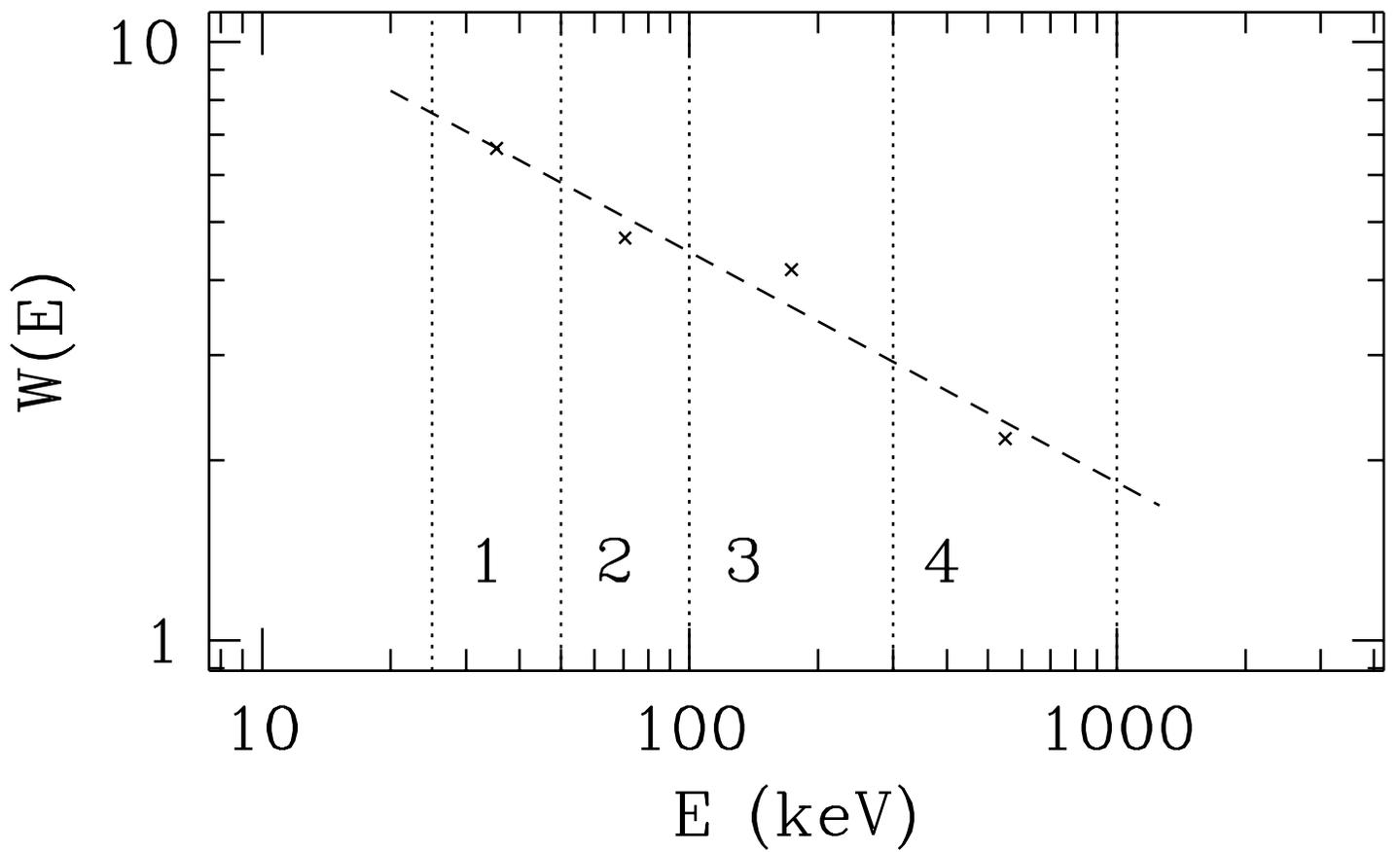
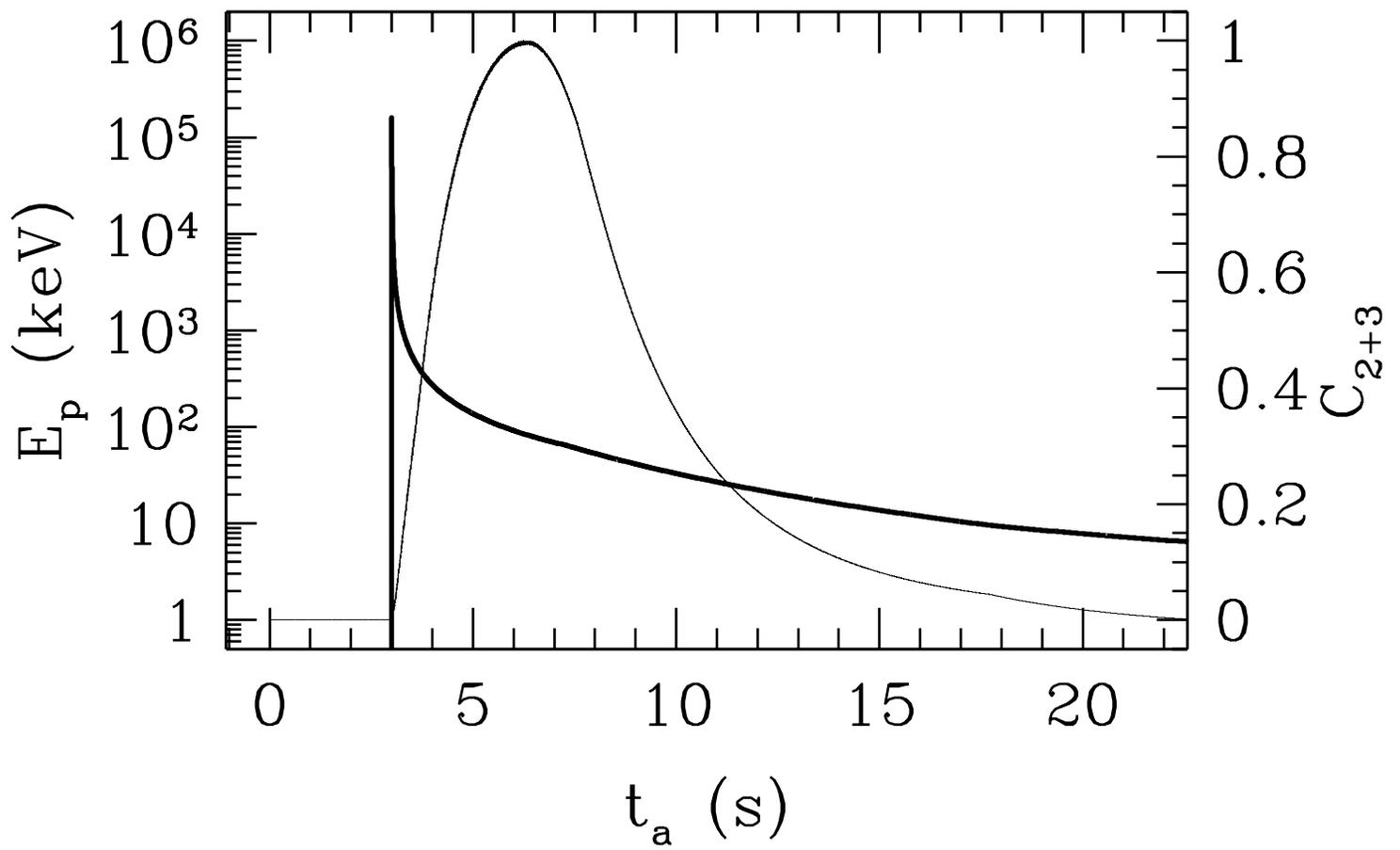
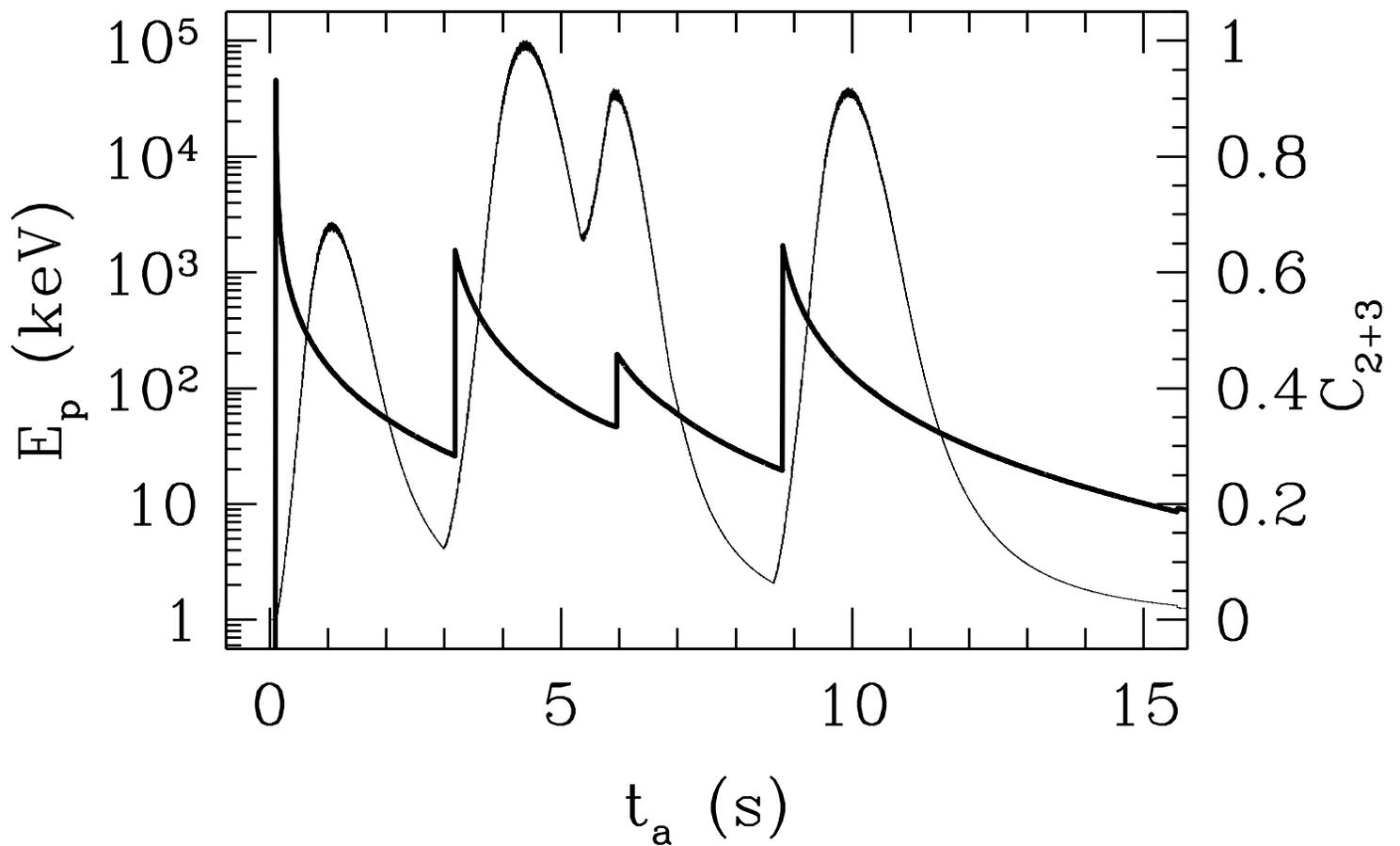
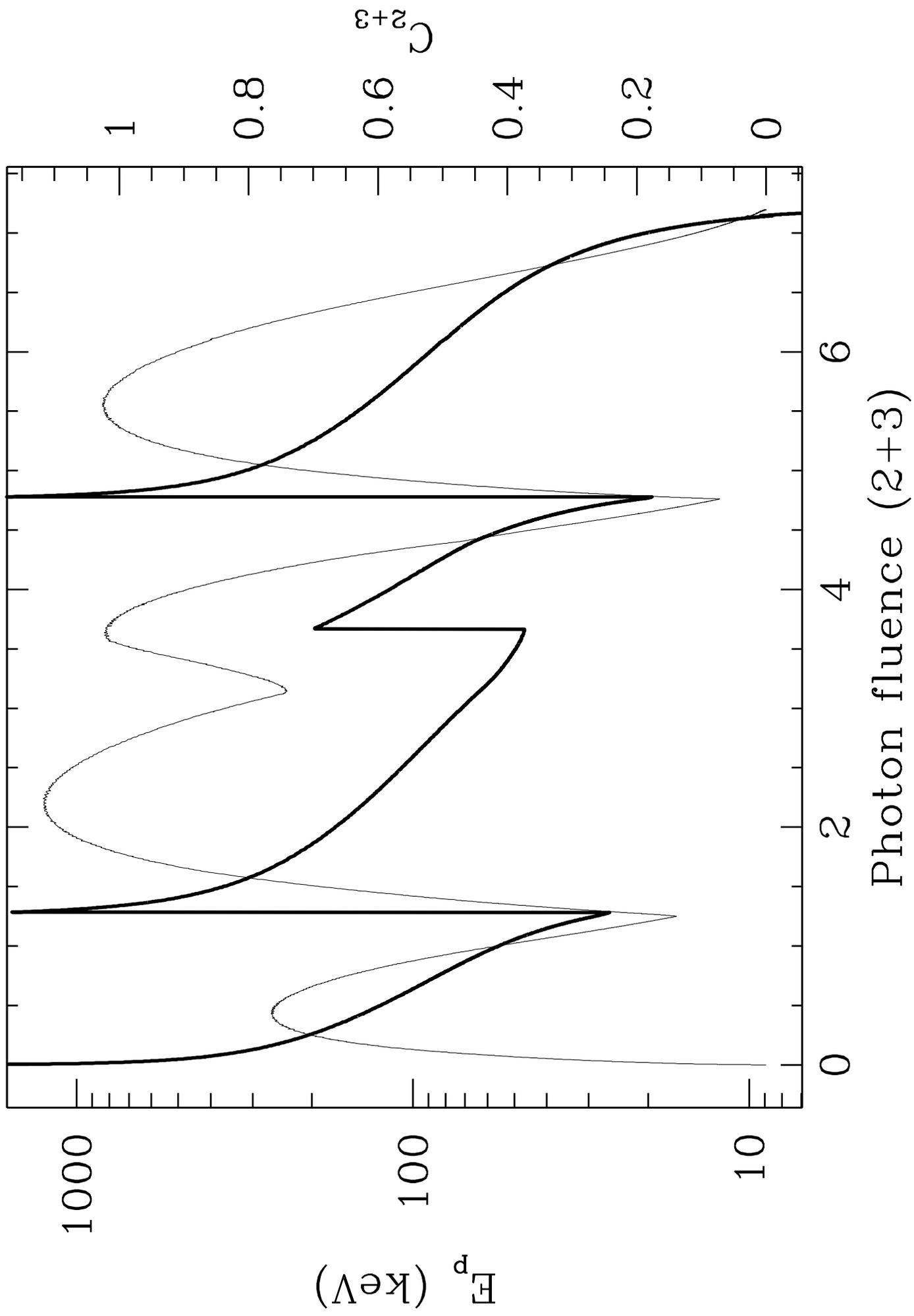
\end